\documentclass[12pt]{article}
\usepackage{a41}
\usepackage{epsfig}
\begin{document}

\newcommand{\be}{\begin{equation}}
\newcommand{\ee}{\end{equation}}

\newcommand{\cc}{\cite}
\newcommand{\ba}{\begin{eqnarray}}
\newcommand{\ea}{\end{eqnarray}}

\begin{center}
\textbf{ Anomalous nonperturbative quark-gluon chromomagnetic
interaction and spin effects in high energy reactions}

\vspace{3mm}

 \underline{N.S.~Korchagin}$^{1}$ and  N.I.~Kochelev$^{2}$,

\vspace{3mm}

\begin{small}
 {Bogoliubov Laboratory of Theoretical Physics,\\ Joint
Institute for Nuclear Research,\\ Dubna, Moscow region,
141980, Russia} \\
(1) {E-mail: korchagin@theor.jinr.ru} \\
  (2) {E-mail: kochelev@theor.jinr.ru} \\

\end{small}
\end{center}


\begin{abstract}
We disscuss a new nonperturbative  mechanism for spin effects  in
 high energy reactions with hadrons. This mechanism is
based on the existence of a large anomalous quark chromomagnetic
moment (AQCM) induced by the nontrivial topological structure of
QCD vacuum. As an example, we estimate the contribution of this
interaction to the single spin asymmetry (SSA) in the inclusive
pion production in the proton-proton scattering. We show that SSA
induced by AQCM  is large and its value is in  qualitative
agreement with  experimental data.
\end{abstract}

\vspace{7.2mm}

The explanation of the large spin effects in high energy reactions
with hadrons is one of the long-standing problems in QCD (see
\cite{Krisch:2010hr,Leader:2001gr} and references therein). It is
well known that
 QCD has a complicated structure of vacuum which leads to the
 phenomenon  of spontaneous  chiral symmetry breaking (SCSB) in
 strong interaction. The instanton liquid model of QCD vacuum
 \cite{shuryak}, \cite{diakonov} is one of the models in which the
 SCSB phenomenon arises  in a very natural way due to the quark
 chirality-flip in the field of strong fluctuation of  vacuum gluon
 fields
 called instantons.
As the result, instantons lead to the anomalous
quark-gluon chromomagnetic  vertex with a large
  quark spin-flip  \cite{kochelev1}\footnote{The importance of SCSB
phenomenon  for quark spin-flip was also mentioned  in the recent
paper \cite{Troshin:2012fr} in the different aspect.}.

Therefore, they can give an important contribution to the spin
dependent cross sections.

 In this Letter, we discuss the mechanism
of spin effects based on the quark spin-flip by the
nonperturbative contribution coming from AQCM. As an example, we
present the estimation of the AQCM contribution to SSA in the
inclusive pion production in the  high energy proton-proton
interaction \footnote{The details of  calculation of the AQCM
contribution to  SSA at the quark level can be found in \cite{
Kochelev:2013zoa}.}.

The general quark-gluon vertex with the AQCM contribution is
\begin{equation}
V_\mu(q^2)t^a = -g_st^a( \gamma_\mu
 +
\frac{\mu_aF_2(q^2)}{2M_q}\sigma_{\mu\nu}q_\nu),
 \label{vertex}
 \end{equation}
where the first term is  the conventional pQCD quark-gluon vertex
and the second term in our  model  comes from the nonperturbative
sector of QCD. In Eq.\ref{vertex}, the form factor $F_2$
describesw nonlocality of the interaction, $\mu_a$ is AQCM, $q^2$
is the virtuality of  the gluon and
 $M_q$ is the dynamical quark mass.

The form factor  $F_2(q^2)$ suppresses the AQCM vertex at short
distances when the respective virtuality is large. Within the
instanton model \cite{shuryak,diakonov} it has the following form
\begin{equation}
F_2(q^2) = F_g(\mid q\mid\rho),
 \nonumber \end{equation}
 where
\begin{equation}
 F_g(z)=\frac{4}{z^2}-2K_2(z)  \label{ffg}
\end{equation}
is the instanton form factor,  $K_2(z)$ is the modified Bessel
function and $\rho$ is the instanton size.

In this model
 AQCM is \cite{kochelev2}
\begin{equation}
\mu_a=-\frac{3\pi (M_q\rho_c)^2}{4\alpha_s}, \label{AQCM}
\end{equation}
where $\rho_c$ is the average size of instantons in  the QCD
vacuum. The value $\mu_a$ of AQCM strongly depends on the
dynamical quark mass $M_q$ generated by SCBS. In the mean field
approximation (MFA)\cite{shuryak}, $M_q=170$ MeV and from
Eq.\ref{AQCM}  ${\mu_a}^{MFA}=-0.4$. In the Diakonov-Petrov model
(DP)\cite{diakonov}, $M_q=350$ MeV and $\mu_a^{DP}= -1.6$. The
strength of nonperturbative interaction in Eq.\ref{vertex} has the
following dependence on $M_q$ and the strong coupling constant
$g_s$
\begin{equation}
V^{nonpert}\sim \frac{M_q}{g_s},
\nonumber
\end{equation}
which clearly shows the relation to the SCSB phenomenon induced by
nonperturbative QCD dynamics.

The SSA for the process  of transversely polarized  quark
scattering  on an unpolarized quark,
 $q^{\uparrow}(p_1)+q(p_2) \to q (p_1^\prime)
+ q(p_2^\prime)$, is defined as
\begin{equation}
A_{N}=\frac{d \sigma^{\uparrow} - d \sigma^{\downarrow}} {d
\sigma^{\uparrow}+d \sigma^{\downarrow}},
\end{equation}
where   ${\uparrow}  {\downarrow}$ denote the initial quark spin
orientation perpendicular to the scattering plane. On the other
hand, the value of SSA can be expressed in terms of the helicity
amplitudes:
\begin{equation}
A_N=-\frac{2Im[( \Phi_1+\Phi_2+\Phi_3-\Phi_4)\Phi_5^*]}
{|\Phi_1|^2+|\Phi_2|^2+|\Phi_3|^2+|\Phi_4|^2+4|\Phi_5|^2)}.
\label{helicity}
\end{equation}
\begin{equation}
\Phi_1=M_{++;++},\ \   \Phi_2=M_{++;--},\ \   \Phi_3=M_{+-;+-},\ \
\Phi_4=M_{+-;-+}  ,\ \   \Phi_5=M_{++;+-}, \nonumber
\end{equation}
where the symbols $+$ or $-$ denote the helicity of a quark in the
c.m. frame.

\begin{figure}[h]
\vspace*{0.5cm}
  \centering 
  \vspace*{-5mm} 
  \includegraphics[width=80mm]{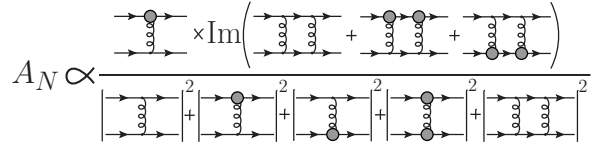}
  \caption{\footnotesize
The contribution to the SSA arising from different diagrams.}
\end{figure}

In  Fig.1, we present the set of   diagrams which give a leading
contribution to $A_N$.
\begin{figure}[h]
\begin{minipage}[c]{8cm}
\vskip -0.5cm \hspace*{-0.5cm}
\centerline{\epsfig{file=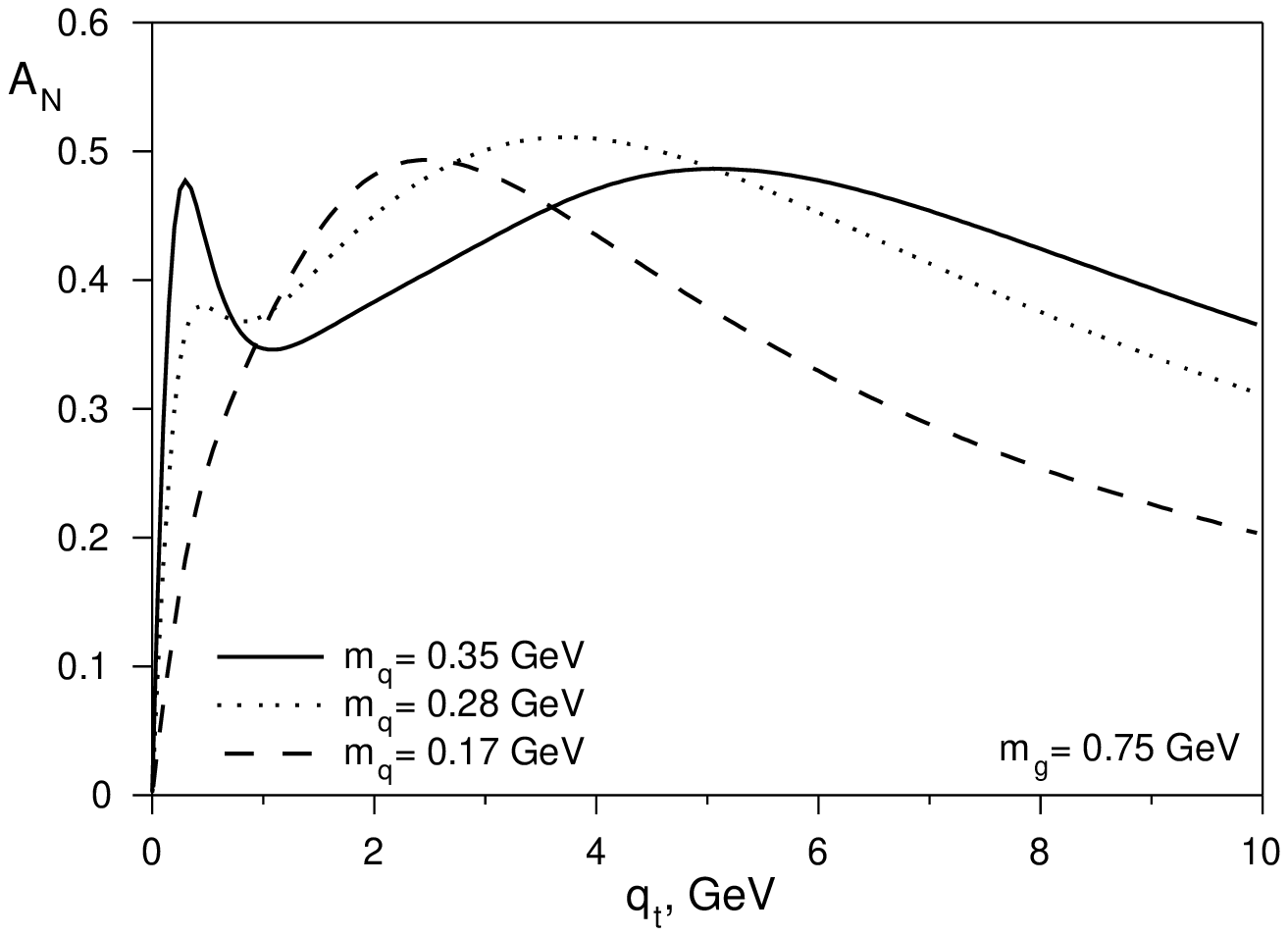,width=9cm}}
\end{minipage}
\begin{minipage}[c]{8cm}
\centerline{\epsfig{file=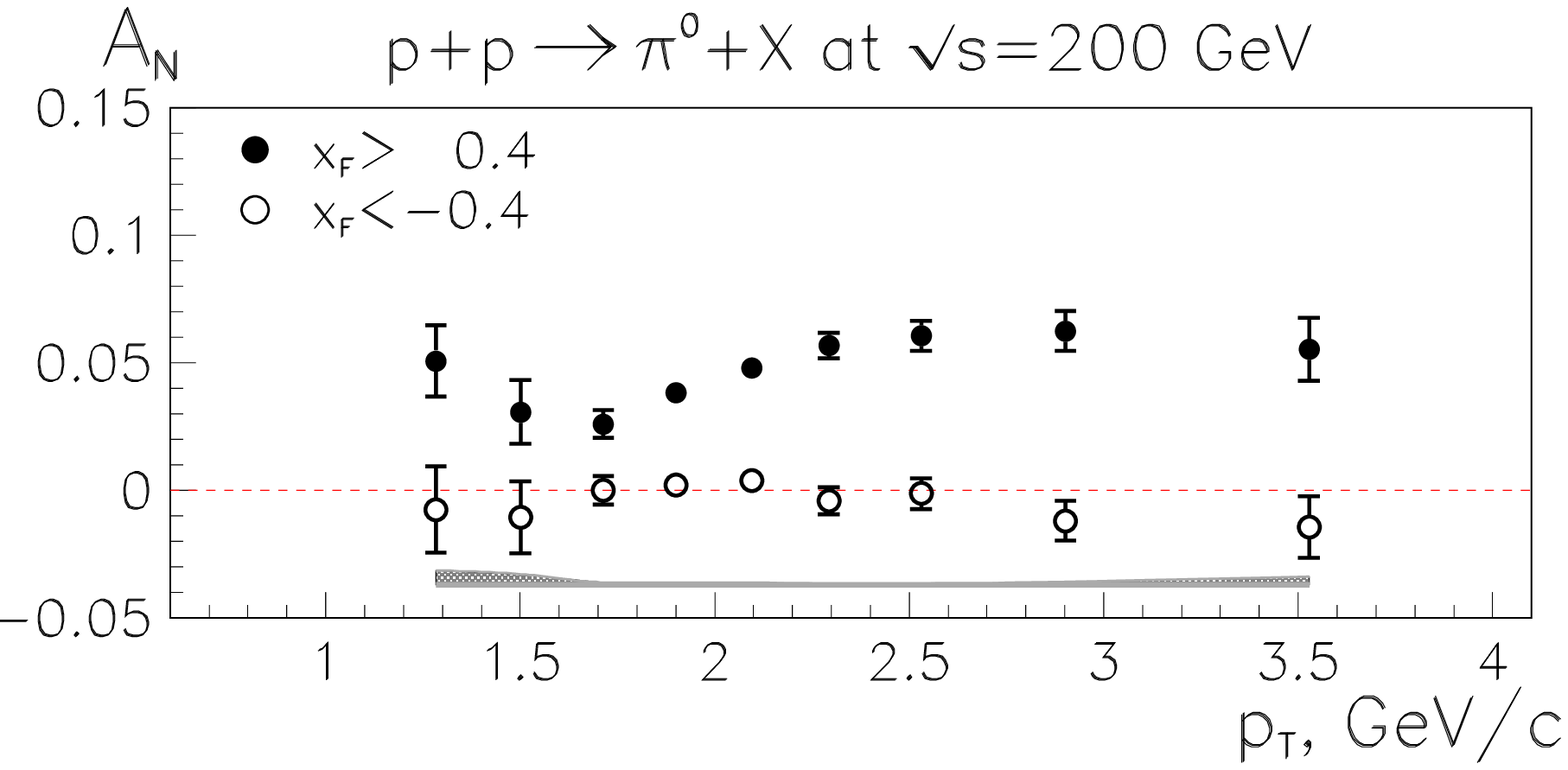,width=8cm}}\
\hspace*{1.0cm} \vskip -1cm
\end{minipage}
\caption{ \footnotesize Left panel: the $q_t$ dependence of  SSA
for the different values of the dynamical quark mass  and fixed
value for the dynamical gluon mass $m_g=0.75$ GeV. Right panel:
STAR data for inclusive $\pi^0$ production \cite{STARpi0:2008}.}
\end{figure}
 The result of the calculation is shown in the
 the left panel of Fig.2. It is evident that $A_N$ induced by AQCM is very large and
       has a rather weak dependence on  $M_q$. We would like to emphasize  that $A_N$ in our approach
        does not depend on  energy. This behavior is directly related to the spin one t-channel gluon exchange.
         This phenomenon is in agreement with experimental data.
 Another remarkable feature of our approach is a  flat dependence of SSA on
 the transverse momentum. It comes from a rather soft power-like
 form factor in the quark-gluon vertex,   Eq.\ref{ffg},
 and a small average size of an instanton, $\rho_c\approx 1/3$ fm \cite{shuryak}.
 Such a flat dependence has been observed by the STAR collaboration in the inclusive $\pi^0$ production
 in high energy proton-proton collision, right panel of Fig.2, and
 was not expected in the models based on the so-called TMD factorization  \cite{collins}.
 Finally, the sign of the SSA is defined by the sign of AQCM and should be positive.
 This sign is very important for explanation of the signs of SSA observed for the inclusive production
 of $\pi^+,\pi^-$ and $\pi^0 $ mesons in proton-proton and proton-antiproton high energy collisions.
 We can estimate asymmetry at the hadron level  for the inclusive production of pions
 in the proton-proton scattering by using some simple
 assumptions. Let us consider only leading fragmentation of pions
 from the final quark. In this case,  SSA for the different charge
 of pions is
 \begin{equation}
 A_N^\pi (q_t)\approx \frac{{\Delta q}_t}{q}A_N^q(q_t),
 \label{hadron}
 \end{equation}
 where $A_N^q(q_t)$ is SSA at the quark level presented in the left panel in Fig.2, ${\Delta q}_t$ is
 the transverse  polarization of the  quark with the given flavor  in the transversely polarized
 proton and $q$ is the number of the corresponding quark   in the
 proton.  Using the additional assumption     ${\Delta q}_t\approx \Delta q $,
 where    $\Delta q$ is the longitudinal polarization of the quark
 in the
 longitudinally polarized proton we have got
 \begin{equation}
 A_N^{\pi^+}(q_t)\approx  0.383A_N^q(q_t),\  \   A_N^{\pi^-}(q_t)\approx
-0.327A_N^q(q_t), \  \ A_N^{\pi^0}(q_t)\approx   0.146 A_N^q(q_t),
 \label{SSAh}
 \end{equation}
 where we used values $\Delta u=0.766$ and $\Delta d=-0.327$ from
 \cite{deFlorian:2009vb}.
 Finally, one can verify that our estimation given by  Eq.\ref{SSAh}  is in  qualitative
 agreement with the available experimental data
 \cite{STARpi0:2008,Lee:2007zzh,  Adams:1991cs} for the large $x_F$ region.

In summary, we discussed the spin effects in  high energy
reactions induced by AQCM. This phenomenon appears from the
anomalous strong spin-flip  quark-gluon interaction induced by the
topologically nontrivial configuration of the vacuum gluon fields
called instantons.  As an example, we estimated the contribution
of AQCM to SSA in the inclusive production of the pions in the
proton-proton scattering and showed that it was large. Additional
arguments for the importance of AQCM for spin effects in high
energy reactions can be found in \cite{ Kochelev:2013csa} where
its contribution to the elastic proton-proton scattering at large
momentum transfer was considered. We would like to mention that
the mechanism of spin effects based on AQCM is quite general and
might happen in any nonperturbative QCD model with  SCSB. The
attractive feature of the instanton model is that within this
model  this phenomenon comes from rather small distances
$\rho_c\approx 0.3$ fm.  As the result, it  allows one to
understand the origin  of large observed
 spin effects at  large transverse momenta.

The authors are very grateful to   A.V. Efremov  for the
invitation to DSPIN-13 Workshop and  for the discussion.

\end{document}